%% file: 0_Main.tex
\documentclass[manuscript,screen]{acmart}
\AtBeginDocument{%
  }

\acmConference[Conference acronym 'XX]{Make sure to enter the correct
  conference title from your rights confirmation email}{June 03--05,
  2018}{Woodstock, NY}
\acmISBN{978-1-4503-XXXX-X/2018/06}

\usepackage{graphicx} 
\usepackage{subfig}  
\usepackage{float}    
\usepackage{multirow} 
\usepackage{makecell}
\usepackage{booktabs}

\usepackage{xcolor}
\newcommand{\ours}{\textit{SoulSeek}}
\newcommand{\interviewquote}[1]{\textcolor{black}{\textit{#1}}}

\begin{document}

\title{SoulSeek: Exploring the Use of Social Cues in LLM-based Information Seeking}


\author{Yubo Shu}
\affiliation{%
  \institution{Fudan University}
  \city{Shanghai}
  \state{Shanghai}
  \country{China}
}
\email{ybshu20@fudan.edu.cn}

\author{Peng Zhang}
\affiliation{%
  \institution{Fudan University}
  \city{Shanghai}
  \country{China}
}
\email{zhangpeng_@fudan.edu.cn}

\author{Meng Wu}
\affiliation{%
  \institution{Lenovo Group}
  \city{Beijing}
  \country{China}
}
\email{wumeng5@lenovo.com}

\author{Yan Chen}
\affiliation{%
  \institution{Virginia Tech}
  \city{Blacksburg}
  \state{Virginia}
  \country{USA}
}
\email{ych@vt.edu}

\author{Haoxuan Zhou}
\affiliation{%
  \institution{Fudan University}
  \city{Shanghai}
  \country{China}
}
\email{24210240428@m.fudan.edu.cn}

\author{Guanming Liu}
\affiliation{%
  \institution{Fudan University}
  \city{Shanghai}
  \country{China}
}
\email{gmliu24@m.fudan.edu.cn}

\author{Yu Zhang}
\affiliation{%
  \institution{Lenovo Group}
  \city{Beijing}
  \country{China}
}
\email{zhangyu29@lenovo.com}

\author{Liuxin Zhang}
\affiliation{%
  \institution{Lenovo Group}
  \city{Beijing}
  \country{China}
}
\email{zhanglx2@lenovo.com}

\author{Qianying Wang}
\affiliation{%
  \institution{Lenovo Group}
  \city{Beijing}
  \country{China}
}
\email{wangqya@lenovo.com}

\author{Tun Lu}
\affiliation{%
  \institution{Fudan University}
  \city{Shanghai}
  \state{Shanghai}
  \country{China}
}
\email{lutun@fudan.edu.cn}

\author{Ning Gu}
\affiliation{%
  \institution{Fudan University}
  \city{Shanghai}
  \state{Shanghai}
  \country{China}
}
\email{ninggu@fudan.edu.cn}

\renewcommand{\shortauthors}{Yubo Shu et al.}


\begin{abstract}
Social cues, which convey others' presence, behaviors, or identities, play a crucial role in human information seeking by helping individuals judge relevance and trustworthiness. However, existing LLM-based search systems primarily rely on semantic features, creating a misalignment with the socialized cognition underlying natural information seeking. To address this gap, we explore how the integration of social cues into LLM-based search influences users' perceptions, experiences, and behaviors. Focusing on social media platforms that are beginning to adopt LLM-based search, we integrate design workshops, the implementation of the prototype system (SoulSeek), a between-subjects study, and mixed-method analyses to examine both outcome- and process-level findings. The workshop informs the prototype’s cue-integrated design. The study shows that social cues improve perceived outcomes and experiences, promote reflective information behaviors, and reveal limits of current LLM-based search. We propose design implications emphasizing better social-knowledge understanding, personalized cue settings, and controllable interactions.
\end{abstract}


\keywords{Social Cues, Social Content Platform, LLM-based Search, Information Seeking}

\maketitle

\input{1_Intro}

\input{2_Related_Work}

\input{3_Design_Workshop_and_Implementation}

\input{4_Empirical_Study_and_Findings}

\input{5_Discussion}

\input{6_Limitation}

\input{7_Conclusion}

\begin{acks}
This work is sponsored by CAAl-Lenovo Blue SkyResearch Fund. Tun Lu is also a faculty of Shanghai Key Laboratory of Data Science, Fudan Institute on Aging, MOE Laboratory for National Development and Intelligent Governance, and Shanghai Institute of Intelligent Electronics Systems, Fudan University.
\end{acks}

\bibliographystyle{ACM-Reference-Format}
\bibliography{8_References}

\end{document}

%% file: 1_Intro.tex
\section{Introduction}

Information seeking is usually viewed as a process in which individuals search for and use information when they encounter knowledge gaps. In the information age, information seeking has become ubiquitous\cite{case2016looking}, and the rapid development of information systems has vastly increased the availability of information while also intensifying information overload\cite{roetzel2019information}. As a result, individuals increasingly rely on multiple cues to filter and make sense of information \cite{bennett2023does, fan2025user, wilson2006revisiting}.
These cues include both message cues (semantic feature in the content)\cite{meyers1991gender} and \textbf{social cues} beyond semantics \cite{winter2016selective}. Social cues can reflect others’ presence, behaviors, or identities in information interactions, such as the identity of the information publisher, the author’s style in the content, or feedback from other readers. Social cues play a key role in human information seeking. 
From the user perspective, social cues serve as contextual factors that enable individuals to more effectively assess information credibility and relevance, especially in situations of information overload and uncertainty, thereby shaping attention allocation and decision-making tendencies \cite{traberg2024persuasive, kosova2025social}. 
From the system design perspective, social cues can be extracted and modeled by information systems to optimize retrieval and recommendation processes, improving both the accuracy of information presentation and the efficiency of system responses \cite{klein2025effects}.

\begin{figure}[h!]
  \centering
  \includegraphics[width=0.55\textwidth]{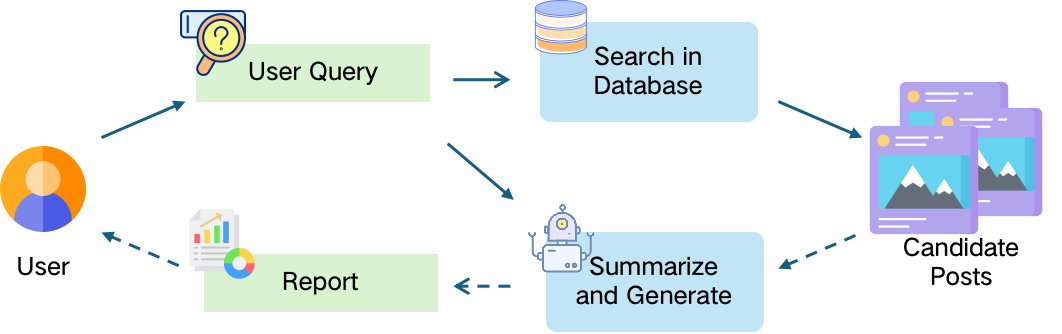}
  \caption{The core pipeline of an LLM-based search system.}
  \label{fig:llm_search_workflow}
\end{figure}

Information, primarily user-generated content (UGC) in social media, usually contains rich social cues\cite{xu2024mental, tian2015combining, wanichayapong2015improving}. In recent years, \textbf{social media platforms} have gradually become important channels for information seeking\cite{mulansky2022social}. For example, users seek information about local events on Facebook \cite{hu2013whoo}, or search for strategies for mental health issues on TikTok \cite{milton2024seeking}. Social media provides not only rich message cues, but also accessible and diverse social cues, such as the poster's public profile, number of followers, and interaction in the comment section. Users can use these cues to support their information seeking activities and help them filter and judge diverse posts. With the rapid development of artificial intelligence technologies, many mainstream social media platforms (e.g., X, Weibo, RedNote) have started integrating \textbf{LLM-based search systems} to assist users' information seeking \cite{gao2023retrieval}. LLM-based search usually follows a core pipeline of ``retrieval–understanding–generation'' \cite{fan2024survey}. As shown in Fig.~\ref{fig:llm_search_workflow}, the pipeline first performs an initial retrieval from massive posts based on the user's query; then the LLM filters, understands, and integrates the retrieved content; finally, it generates a natural language response. LLM-based search systems can understand natural language queries, match multi-source information, and automatically organize and generate comprehensive answers, thereby significantly improving the efficiency of information seeking\cite{gao2024easyask}.

With limited integration and utilization of social cues, existing LLM-based search systems still primarily rely on semantic cues, which \textbf{ misalign
 to the cognitive socialization of human during natural information seeking}. Namely, the limitations are mainly reflected in that the systems usually only filter and summarize information based on semantic computation or rule-based matching (e.g., keyword matching) \cite{wang2025social}, without fully considering the user's need for social cues beyond semantics. For instance, when a user searches for posts, they might want the system to prioritize content with positive feedback in the comment section or high discussion popularity, which reflects social approval or group consensus. However, this type of social cue-based retrieval intent is currently not supported by mainstream LLM-based search tools, thereby limiting their effectiveness in socialized information seeking scenarios.

Although the aforementioned misalignment has been noted, as an emerging paradigm of information retrieval, \textbf{integration of social cues and its mechanism in LLM-based search still lacks systematic research}, it remains unclear how users understand and expect social cues to be presented in the results before searching, how they control and filter results based on social cues during the search process, and how they interpret and utilize results with social cues after searching. To address this gap, we conduct an exploratory study for the following research questions: \textbf{RQ1: In the context of LLM-based information seeking, what social cues do users expect to use, and how should these social cues be used? RQ2: How can the integration of social cues affect users’ information seeking outcome and process?}

To conduct this study, we selected LLM-based search tools provided by social media as the research context. Social media platforms have not only become an important channel for people to seek information but also provide rich and easily accessible social cues. Furthermore, several mainstream platforms have recently introduced or are experimenting with LLM-based search, further demonstrating the breadth and practical significance of this trend. Based on these factors, social content platforms offer a representative context for studying the combination of social cues and LLM-based search. We adopted an exploratory research methodology, with core processes including prototype deployment and empirical study. For RQ1, we organized workshops to collect user needs and expectations for social cues, and accordingly designed and implemented an LLM-based search prototype system, \ours, which supports social cue integration. For RQ2, we recruited users to use this prototype in a real information seeking context, and through a between-subjects experiment, recorded their operational and interaction processes. Finally, we adopted a mixed-method approach that combined qualitative and quantitative analyses to systematically derive research findings at both the outcome and process levels. The results of this study reveal the multifaceted role of social cues in LLM-based search: the integration of social cues significantly improved users' perception and experience, enhanced the usefulness and trustworthiness of the results, made the search process more directional and controllable, and prompted users to reflect on their information behavior. At the same time, the study found that the model still has limitations in understanding socialized knowledge and balancing multi-cue matching. This paper further points out that the active use of social cues by users helps to restore and strengthen their agency in information seeking, and accordingly proposes design implications for future social cue-aware LLM search systems, including enhancing social knowledge understanding, supporting personalized cue configuration, and realizing a transparent and controllable human-computer collaborative search process. In summary, the contributions of this paper include:


\begin{itemize}
\item This work is among the first to investigate the mechanisms of social cues in LLM-based search.
\item We conducted design workshops to identify user needs and developed a prototype system \ours, which supports social cue integration.
\item Through empirical studies, we revealed how social cues shape users’ perceptions, experiences, and behaviors.
\item We derive design implications for future LLM-based search, guiding the integration of social cues into design.
\end{itemize}


%% file: 2_Related_Work.tex
\section{Related Work}

\subsection{Information seeking}

Information seeking is a fundamental human cognitive and social process in which individuals recognize an information need and actively collect, filter, and integrate information to support understanding and decision-making \cite{wilson1999models, pirolli1995information, kuhlthau1991inside}. In complex contexts such as academic research or investigative work, it is highly iterative and non-linear, involving continuous interaction under uncertainty and integration of multi-source information \cite{pirolli2009powers}.

Theories on information seeking have evolved from efficiency-oriented to experience-oriented paradigms. Information Foraging Theory emphasizes strategic cost–benefit trade-offs in information environments \cite{pirolli1999information}, while Sense-making Theory highlights the contextual cycles through which users interpret and organize information \cite{dervin1983overview}. The notion of the Information Flaneur further frames seeking as an exploratory and reflective practice \cite{dork2011information}. Together, these theories reveal the dynamic and experiential nature of information seeking. Building on these foundations, HCI research has examined information seeking across diverse platforms and contexts. Studies have shown how social contexts shape users’ health information behaviors \cite{de2014seeking}, how community cues foster belonging in hyperlocal search \cite{hu2013whoo}, and how personal information ecosystems support iterative verification and reflection \cite{milton2024seeking}. Across these works, users are understood not as passive receivers but as active shapers of information pathways and meanings \cite{bennett2023understand, fan2025user}. Recognizing user agency has thus become central to designing more personalized and adaptive information systems.

This study builds on this literature by focusing on social content platforms, which embed rich social cues such as author identity, feedback, and conversational tone. These cues shape users’ judgments and sense-making, making them essential for understanding the socialized nature of information seeking and for designing more context-aware intelligent systems \cite{kim2015use, milton2024seeking}.

\subsection{Social cues in information seeking}

Social cues are signals that convey social or interpersonal information and influence how people interpret and respond to others \cite{winter2016selective, pennycook2023social,shu2024understanding}. In face-to-face communication, they include nonverbal behaviors such as gaze, expression, and tone, while on social platforms they appear as likes, comments, author identity, and engagement levels. Feine et al. \cite{feine2019taxonomy} proposed a taxonomy of social cues comprising four types—verbal, visual, auditory, and invisible—capturing both explicit and implicit features from language to timing.

Social cues play a critical role in information seeking. Winter et al. \cite{winter2016selective} showed that information selection depends not only on content but also on users’ social motivations, with cues like popularity often processed subconsciously. From the Computer-Mediated Communication perspective, Tanis and Postmes \cite{tanis2003social} found that cue availability shapes impression formation, while Huang et al. \cite{huang2024signaling} and Liu et al. \cite{liu2024socialcueswitch} revealed that users flexibly adapt or switch cues across contexts and modalities to convey intent and enhance accessibility. 
Furthermore, Yang et al. \cite{yang2025socialmind} show that users exhibit social-emotional responses, such as admiration and anxiety, when reading social media content, which can be accurately detected by sensor-based measurements. Collectively, these studies highlight the pervasive influence of social cues in shaping attention, trust, and interaction patterns. Recent work also integrates social cues into intelligent systems. Li et al. \cite{li2024context} demonstrated that combining social relationships with visual features improves retrieval performance. Klein’s meta-analysis \cite{klein2025effects} found that human-like cues modestly increase trust and satisfaction, though excessive anthropomorphism may trigger the uncanny valley effect \cite{ciechanowski2019shades}. Reinkemeier et al. \cite{reinkemeier2021voice} further noted that cue effects are highly context-dependent, varying with user expectations and task settings.

In summary, prior research has established the importance of social cues in information seeking, yet most studies emphasize their presentation and perception rather than how users actively employ and configure them—particularly in emerging LLM-based interactions.

\subsection{LLM-Based Search System}

The rapid development of large language models (LLMs) and retrieval-augmented generation (RAG) has enabled new forms of search that combine retrieval with natural language generation \cite{lewis2020retrieval, asai2024self, fan2024survey}. Users express information needs in natural language, and the LLM parses queries, retrieves relevant knowledge, and generates integrated responses. These systems offer flexible, interactive, and iterative search experiences that allow users to refine goals across conversational turns. Despite their promise, current LLM-based search systems face notable limitations. At the knowledge level, RAG methods excel at factual retrieval but struggle with contextual understanding. Wang et al. \cite{wang2025social} showed that ignoring group interactions and situational context often leads to misaligned outputs. At the interaction level, the retrieval–generation process remains opaque, limiting user understanding and control; visual tools such as RAGTrace \cite{cheng2025ragtrace} have been proposed to enhance transparency and explainability.

Recent studies have thus emphasized user agency and reflectivity. While conversational search simplifies interaction, it can also diminish users’ exploratory initiative. Schroeder et al. \cite{schroeder2025forage} observed that automatic summarization may turn users from active seekers into passive recipients. Mei et al. \cite{mei2025interquest} advocated for mixed-initiative collaboration, where systems and users jointly shape search goals. Similarly, Neshaei et al. \cite{neshaei2025user} argued that systems should promote reflection rather than replace user thinking.

In summary, prior work advances LLM-based search design in two directions: improving transparency and restoring user initiative. Building on this trend, this paper explores how integrating social cues can further support natural and controllable information seeking on social content platforms.

%% file: 3_Design_Workshop_and_Implementation.tex
\section{Design Workshop and Implementation}

\subsection{Design Workshop}

To explore our RQ1 and to inform the system design addressing RQ2, we conducted a two-phase design workshop that examined how users understand, use, and expect social cues when engaging with LLM-based search on social content platforms. The workshops aimed to identify which social cues users rely on during information seeking and how these cues should be integrated into the interaction and generation processes of LLM-based search tools.


We selected RedNote (Xiaohongshu)\footnote{\url{https://www.xiaohongshu.com/explore}} as the research context for the following reasons. As one of the most influential social media platforms in China, RedNote has over 350 million monthly active users and an average daily usage time exceeding 74 minutes. The platform supports diverse usage scenarios, and its users frequently engage in information seeking activities. According to its 2024 annual report, approximately 70\% of active users use the search function for information seeking within the platform, covering domains such as travel, consumption, learning, and career planning. Consequently, RedNote has become an important environment for studying information seeking behaviors.

In total, we recruited 10 active users (aged 25–35, with an equal gender ratio of 5:5, all holding at least a bachelor's degree). All participants had over one years of experience using the platform and reported using the search function more than three times per week to obtain practical information. The workshops were conducted via online meetings, with each session lasting approximately two hours. With informed consent, all sessions were audio-recorded and documented, and each participant received compensation of \$10.

We followed the participatory design paradigm \cite{muller1993participatory} and adopted a practical, iterative two-phase workshop structure.
\textbf{Phase-1:} We asked participants to recall a recent episode on RedNote in which they solved a problem via search (e.g., travel, consumption, job seeking) and to explain the main factors that shaped their content choices. We then provided example cue cards (e.g., ``author avatar,'' ``like count,'' ``comment climate,'' ``location tag''). Participants added missing cues and wrote down elements they considered important. Through card clustering and labeling, groups discussed and formed high-level categories, and voted on the importance of each cue type. The groups clustered and labeled the cards, formed higher-level categories through discussion, and voted on the importance of each cue type. The outcome was a diverse set of social-cue instances and a structured taxonomy that grounded subsequent system design. \textbf{Phase-2:} We reviewed Phase-1 outcomes and presented the major cue types. They provided several scenario sketches of the LLM-based search workflow, illustrating a typical process (the user enters a query, the AI returns results, and displays post-level social information). Participants considered how the system should interpret and use these social cues and annotated where and how cues should be introduced using sticky notes and diagrams. The groups then role-played search dialogues to test the plausibility of the proposed mechanisms. The research team synthesized the discussions and sketches and consolidated the co-created ideas into three key modules.

With participant consent, we audio-recorded all workshops and transcribed them using a speech-recognition tool. The first author verified and corrected the transcripts. We conducted a thematic analysis \cite{braun2006using} and used MAXQDA Tool to systematize coding and examine relationships among codes. Two core researchers independently read all feedback to familiarize themselves with the data and to identify initial concepts. We then refined the coding scheme through iterative cycles: open coding to capture emergent concepts and patterns; axial coding to merge semantically similar initial codes into higher-level themes and remove redundant or weak items; and selective coding in which the research team jointly adjudicated and distilled codes related to two core themes—types of ``social cues and their stage-specific use'', and ``integration approaches and design expectations''. These steps ensured logical consistency and alignment with the research goals.

Overall, users expressed a clear need to actively leverage social cues in conversational search on social content platforms. They also identified gaps in how current systems model and respond to these cues. As shown in Table~\ref{tab:context_expection}, the search process has three stages—pre-search, in-search, and post-search—and social cues arise from three sources: the publisher (From), the content itself (Self), and audience feedback (To). In the pre-search stage, users specify cue-based constraints in their queries and augment them with their own social cues on the platform (e.g., follows, interaction history). In the in-search stage, users inspect multi-dimensional social cues of candidate posts, validate the system's matches, and provide immediate feedback, which helps calibrate cue recognition and ranking.
In the post-search stage, users refine their queries based on the surfaced social characteristics, probe latent preferences, and optimize subsequent searches.

\begin{table}[H]
\centering
\caption{Use expectations for information seeking with social cues, categorized by cue sources and information seeking stages.}
\label{tab:context_expection}
\footnotesize
\renewcommand{\arraystretch}{1.3}
\begin{tabular}{p{2cm} p{3.5cm} p{3.5cm} p{3.5cm}}
\toprule

\textbf{Cue source} & \textbf{Pre-search} & \textbf{In-search} & \textbf{Post-search} \\

\midrule

Publisher (From)
& Define query constraints based on publisher features; augment with the user's personal social cues
& Identify and match publisher-side social cues (e.g., profile bio, follower count, tags)
& Based on the matched results, analyze feature associations across publishers \\

\midrule

Content itself (Self)
& Define query constraints based on content features; augment with the user's personal social cues
& Extract post-level features (e.g., location, style, time) and perform matching
& Based on the matched results, analyze feature associations across posts \\

\midrule

Audience feedback (To)
& Define query constraints based on audience feedback features; augment with the user's personal social cues
& Analyze interaction features (e.g., likes, saves, comments) and perform matching
& Based on the matched results, analyze cross-post associations between audience metrics and comment content \\

\bottomrule
\end{tabular}
\end{table}

\begin{figure}[h!]
\centering

\begin{minipage}[t]{0.47\textwidth}
    \centering
     \subfloat[Extracting social cues of posts and users]{\label{fig:module_extract}
    \includegraphics[width=\textwidth]{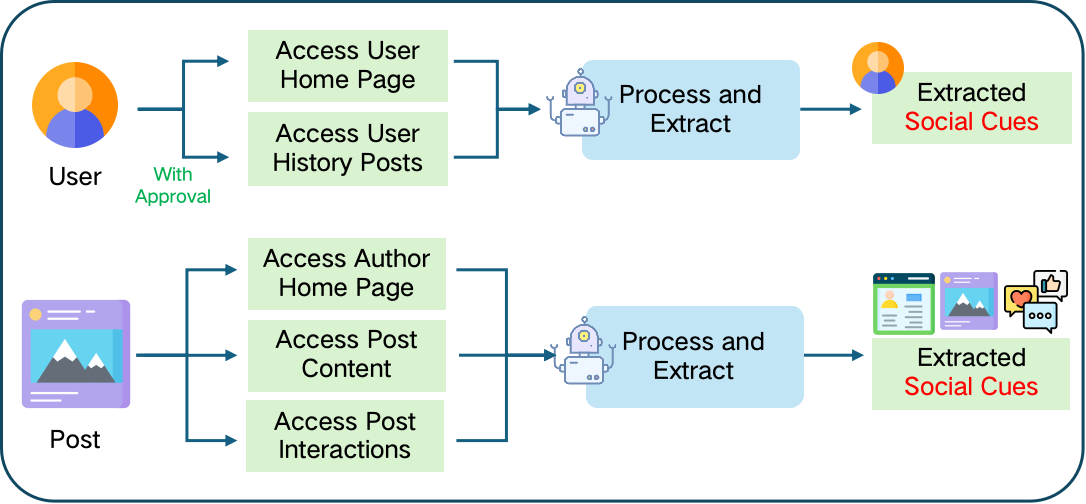}}\\[2ex] 
    \subfloat[Refining user query on social cues]{\label{fig:module_refine}
    \includegraphics[width=\textwidth]{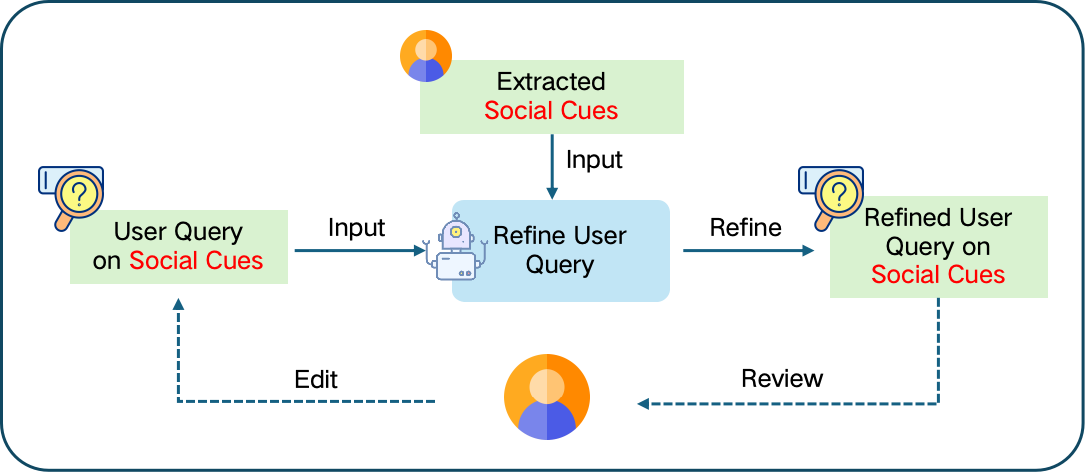}}

\end{minipage}
\hfill
\begin{minipage}[t]{0.49\textwidth}
    \centering
    \subfloat[Matching with social cues]{\label{fig:module_match}
    \includegraphics[width=\textwidth]{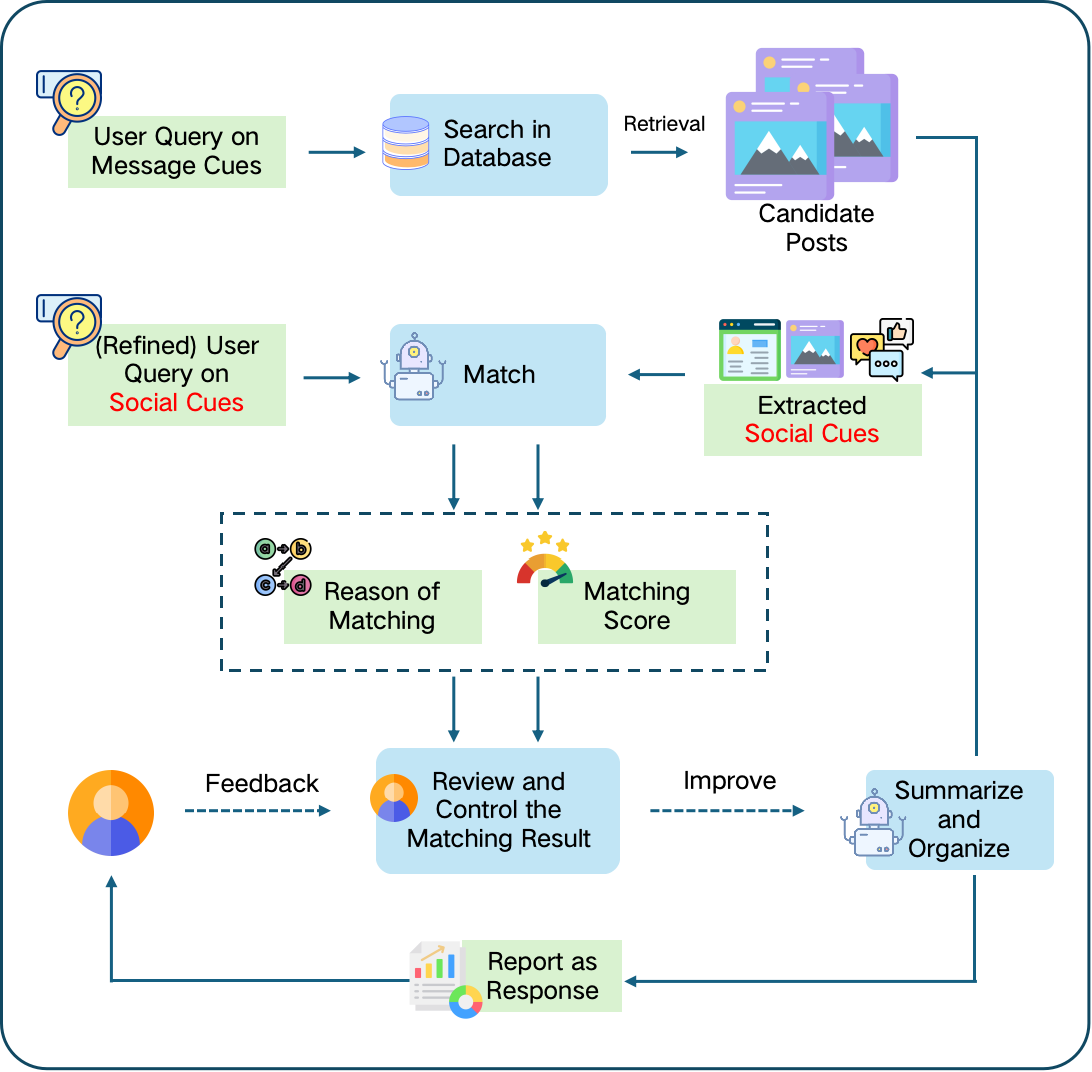}}
\end{minipage}

\caption{Core workflow design for integrating social cues into LLM-based search systems.}
\label{fig:context_aware_modules}
\end{figure}

In the second workshop, we specified how users expect social cues to be introduced and used within the LLM-based search workflow. We summarized these expectations into three key modules shown in Fig.~\ref{fig:context_aware_modules}: (1) Social cue extraction. The system first extracts personal social cues from the user's profile and posting history. It also extracts multi-source social information from retrieved posts, including author-profile attributes, post attributes, and comment-section interaction signals, yielding a multi-level representation of social cues. (2) Social cue query refinement. After the user provides initial cue conditions, the system leverages the user's social history (e.g., prior posts, interaction preferences) to refine and complete the query, making the cue specification more accurate and personalized. (3) Social cue-aware matching and generation. The system combines the user's content query with cue conditions to retrieve candidate posts from the platform database. An LLM then extracts and matches the posts' social cues, produces a match score with rationales, and supports user inspection and control. Finally, the system generates a synthesized answer and report for information seeking based on the user-confirmed results.

\subsection{The Final Design and Implementation}
\begin{figure}[h!]
  \centering
  \includegraphics[width=0.7\textwidth]{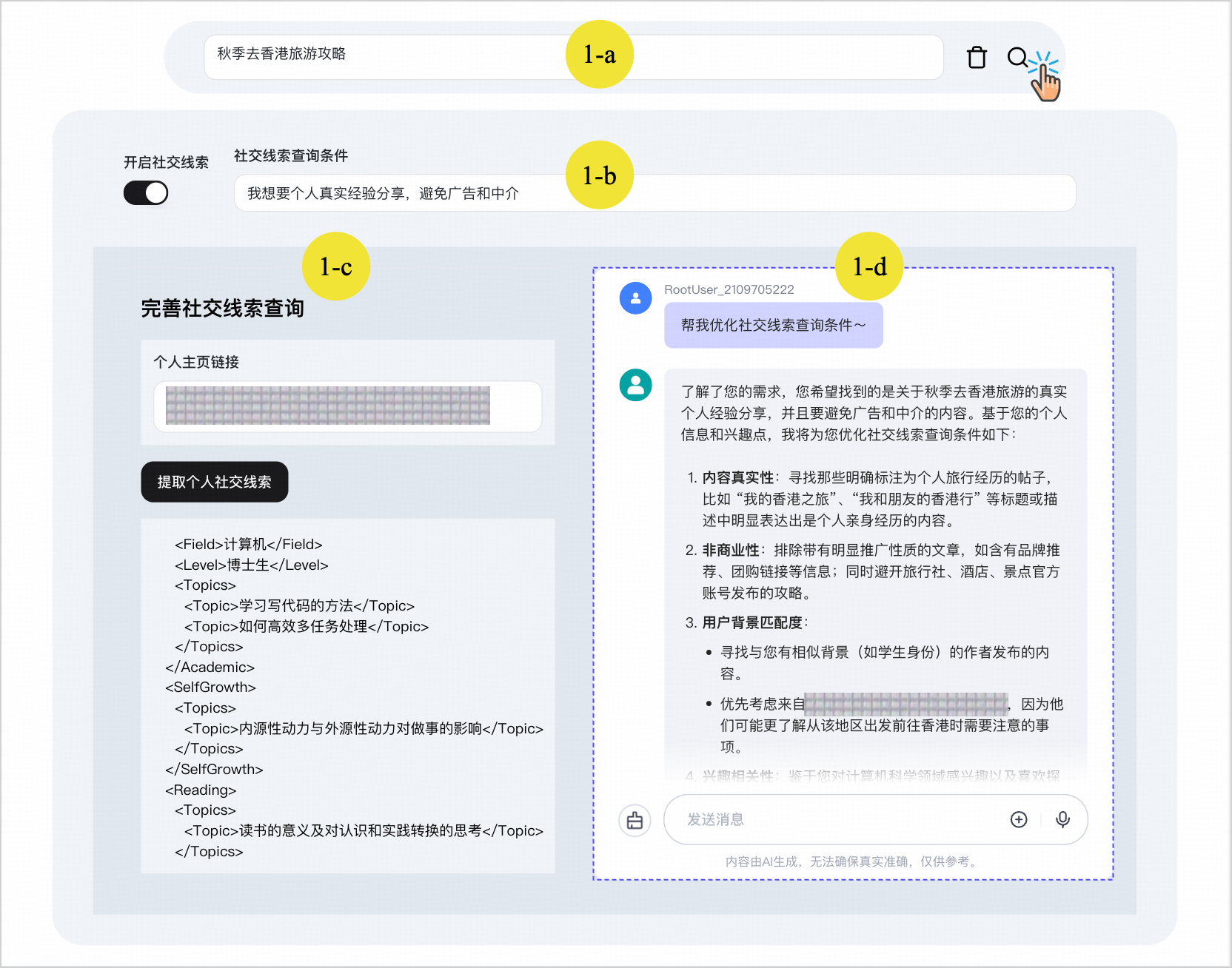}
  \caption{Pre-search user interface in \ours.}
  \label{fig:ui_before_search}
\end{figure}

To validate the effectiveness of social cue integration in real-world settings, we designed and implemented an interactive prototype, \ours, informed by the workshop findings and deployed within a social content platform environment. The interaction flow consists of three stages:

\textbf{In the pre-search stage}, the system adds a social cue input panel alongside the standard search box (Fig.~\ref{fig:ui_before_search}-1-a), allowing users to specify cue preferences and constraints while entering a content query (Fig.~\ref{fig:ui_before_search}-1-b). For example, when searching ``Singapore independent travel guide,'' a user can add ``prefer authentic user experiences; avoid commercial promotions.'' The system parses these cue conditions and forms a multi-dimensional social cue pattern covering the publisher, the content itself, and audience feedback. With user authorization, the system also extracts individual social cues from historical behaviors (Fig.~\ref{fig:ui_before_search}-1-c). Users can then refine their conditions through a dialogue interface with the LLM (Fig.~\ref{fig:ui_before_search}-1-d). Clicking the search icon in the top-right corner of the interface directs users to the in-search stage.

\begin{figure}[h!]
  \centering
  \includegraphics[width=0.7\textwidth]{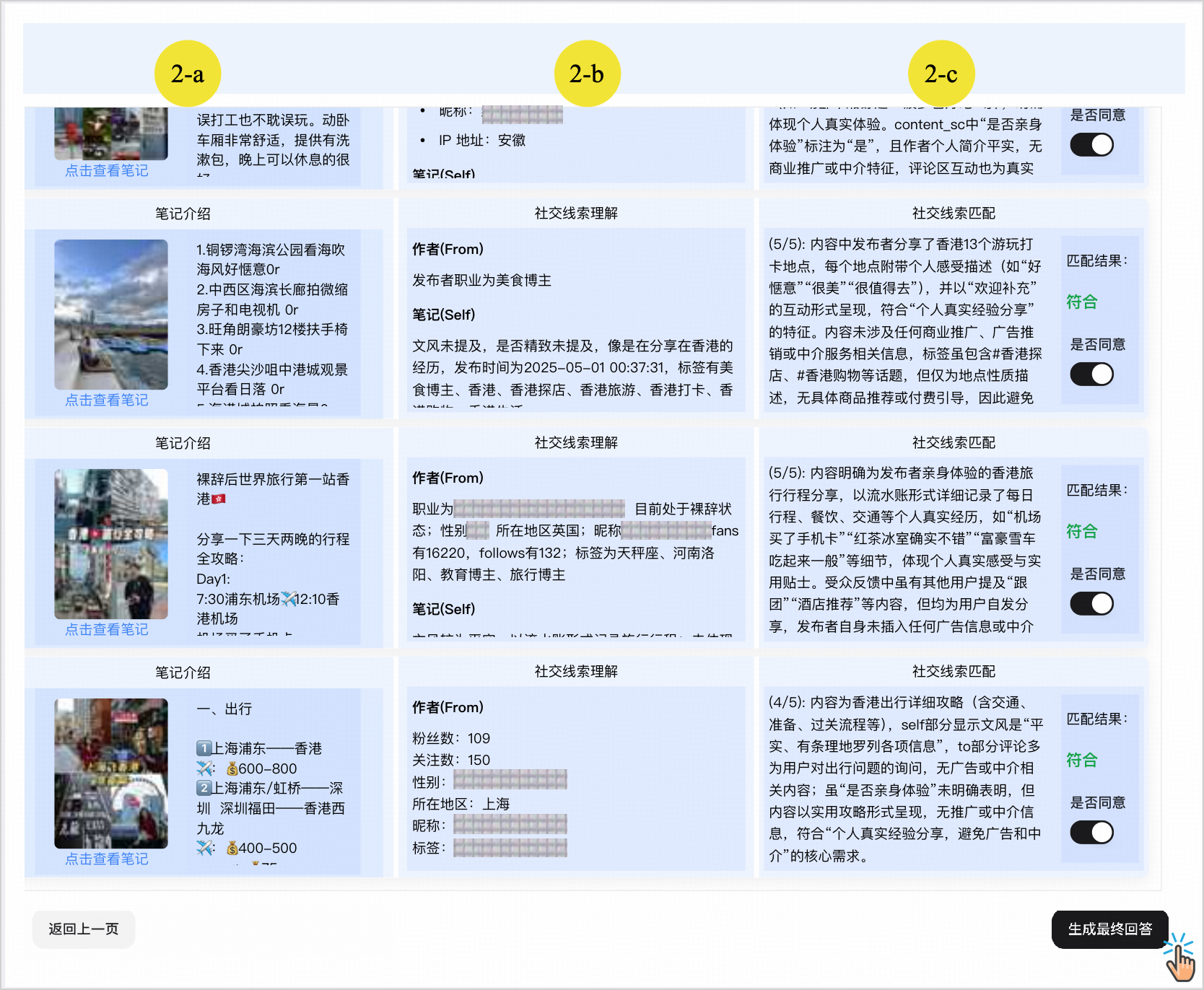}
  \caption{In-search user interface in \ours.}
  \label{fig:ui_during_search}
\end{figure}

\begin{figure}[h!]
  \centering
  \includegraphics[width=0.7\textwidth]{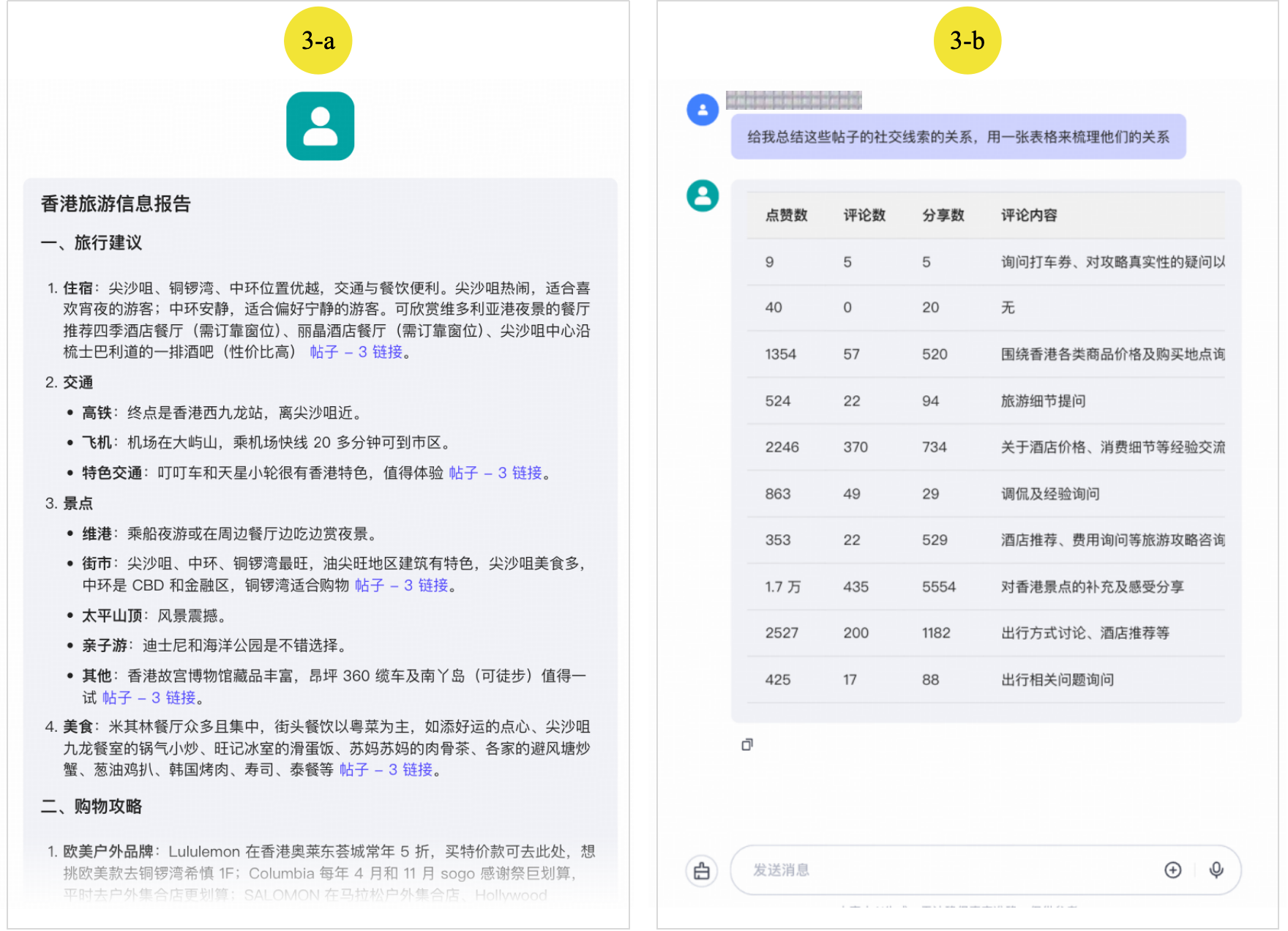}
  \caption{Post-search user interface in \ours.}
  \label{fig:ui_after_search}
\end{figure}

\textbf{In the in-search stage}, the system uses an LLM to extract and match social cues from candidate posts and generates match results with explanations. The left column lists the retrieved posts; clicking a post cover opens the content platform for details (Fig.~\ref{fig:ui_during_search}-2-a). The middle column shows LLM-extracted social cues for each post, covering the publisher, the content itself, and audience feedback (Fig.~\ref{fig:ui_during_search}-2-b). The right column presents the model's match results and rationales based on the user's cue conditions (Fig.~\ref{fig:ui_during_search}-2-c). We use a 1–5 scoring scheme: 5 indicates high relevance to the user's conditions, and 1 indicates no relevance. In our implementation, items scoring above 3 are considered to meet the user's social cue constraints. The system also provides an ``agree with match'' toggle for quick feedback and control. If a model-labeled match does not meet expectations, the user can mark ``disagree.'' The system logs this feedback and, in the subsequent summary stage, feeds it back to the LLM to refine the matching logic and explanations, thereby enhancing user control over the search process. Clicking the ``Generate Report'' button in the bottom-right corner of the interface directs users to the post-search stage.

\textbf{In the post-search stage}, the system compiles the results and generates a summary report (Fig.~\ref{fig:ui_after_search}-3-a). The report highlights items that highly match the user's needs and clearly annotates their sources, helping users understand the provenance and context of the results. Users can continue to interact with the model to explore associations among social cues in the results (Fig.~\ref{fig:ui_after_search}-3-b). For example, they can ask the LLM to synthesize divergent viewpoints from the comment section and extract consensus, enabling a higher-level understanding of the information.

The system uses Qwen-Max as the LLM. It is part of the Qwen family, a leading line of open-source LLMs that combines high performance with local deployability, enabling stable operation in data-private environments. Qwen-Max inherits strong capabilities in general understanding and generation. On the LMArena text leaderboard\footnote{\url{https://lmarena.ai/leaderboard/text}}, it ranks third, outperforming GPT-5-Chat. Given the close coupling between social cues and personal identity, we adopted the Qwen family to ensure applicability in controlled environments that prioritize data security and user privacy, aligning with the emerging trend toward privacy-preserving and edge-deployed AI systems \cite{tuli2025sela}. The prototype is built on the Coze framework, the front end supports search interactions and social cue visualization. The back end use LLM for social cue extraction, query refinement, and result generation. The system strictly adheres to a user-authorization policy when accessing personal information. If a user does not provide a profile link, the system does not automatically extract social cues to refine the query. Users can still manually specify cue conditions. During search, the system accesses only publicly visible content on the platform, consistent with normal usage. All experimental data are anonymized upon storage to protect participant privacy and data security.

%% file: 4_Empirical_Study_and_Findings.tex
\section{Empirical Study and Findings}

\subsection{Study Protocol}

\subsubsection{Between-Subjects Setting}

Building on the previous workshops and the prototype system, we conducted an empirical study with a between-subjects setting to examine the role of social cues in information seeking. The core objective of this study was to address the following research question: How can the integration of social cues affect users’ information seeking outcome and process? To this end, we approached this question from: 
(1) Outcome-level effects—changes in users’ perceptions of search results, such as perceived usefulness and trustworthiness—corresponding to our findings in F1 (\S\ref{sec:f1}); and
(2) Process-level effects—changes in users’ information-seeking behaviors and experiences, including their sense of direction, serendipity, user control, and willingness to use—corresponding to F2 (\S\ref{sec:f2}) and F3 (\S\ref{sec:f3}).

We recruited 20 active RedNote users (10 males and 10 females), all of whom were independent from participants in the previous workshops and held at least a bachelor’s degree. Each participant had over one year of experience using the platform. In addition, all participants were familiar with and had prior experience using both search in RedNote and LLM-based AI search tools, indicating their habitual engagement with information seeking activities on the platform.Detailed demographic information is presented in Table~\ref{tab:participants}. The experiment adopted a between-subjects design, with participants randomly assigned to either the experimental group (E) or the control group (C). Both groups completed the same set of tasks using our developed system. The system used by the experimental group supported the input and presentation of social cues, whereas the control group's system did not include this feature. Aside from the social cue, the two versions of the system were identical in interface and functionality, ensuring the comparability of results.

\begin{table}[htbp]
\centering
\small
\caption{Demographic information of participants in the control and experimental groups.}
\label{tab:participants}
\begin{tabular}{lccccccc}
\toprule
\textbf{ID} &
\textbf{Age} &
\makecell[c]{\textbf{Usage}\\\textbf{of RedNote}\\\textbf{(years)}} &
\makecell[c]{\textbf{Frequency}\\\textbf{of RedNote}\\\textbf{Search (times)}} &
\makecell[c]{\textbf{Frequency}\\\textbf{of LLM-based}\\\textbf{Search (times)}} &
\textbf{Gender} &
\textbf{Education} &
\textbf{Group} \\
\midrule
C1 & 27 & 3 & $>$3/week & $>$3/week & Female & Bachelor & Control \\
C2 & 26 & 4 & $>$3/week & $>$3/day & Male & Master & Control \\
C3 & 25 & 2 & $>$3/day & $>$3/week & Female & Bachelor & Control \\
C4 & 27 & 5 & $>$3/day & $>$3/day & Male & Bachelor & Control \\
C5 & 23 & 3 & $>$3/week & $>$3/month & Female & Bachelor & Control \\
C6 & 30 & 6 & $>$3/day & $>$3/week & Female & Master & Control \\
C7 & 28 & 1 & $>$3/month & $>$3/day & Male & Master & Control \\
C8 & 23 & 4 & $>$3/day & $>$3/month & Male & Bachelor & Control \\
C9 & 25 & 4 & $>$3/day & $>$3/month & Male & Bachelor & Control \\
C10 & 31 & 2 & $>$3/day & $>$3/month & Female & Master & Control \\
\midrule
E1 & 24 & 6 & $>$3/week & $>$3/month & Female & Bachelor & Experimental \\
E2 & 27 & 3 & $>$3/week & $>$3/day & Male & Master & Experimental \\
E3 & 23 & 2 & $>$3/day & $>$3/week & Female & Bachelor & Experimental \\
E4 & 29 & 5 & $>$3/day & $>$3/week & Male & Master & Experimental \\
E5 & 33 & 4 & $>$3/day & $>$3/month & Female & Master & Experimental \\
E6 & 24 & 7 & $>$3/day & $>$3/week & Female & Bachelor & Experimental \\
E7 & 24 & 2 & $>$3/week & $>$3/day & Male & Bachelor & Experimental \\
E8 & 27 & 3 & $>$3/day & $>$3/day & Male & Bachelor & Experimental \\
E9 & 26 & 2 & $>$3/day & $>$3/month & Female & Bachelor & Experimental \\
E10 & 26 & 3 & $>$3/day & $>$3/week & Male & Master & Experimental \\
\bottomrule
\end{tabular}
\end{table}

Considering participants' task load, each experimental session was limited to one hour. Each participant proposed two search queries based on their own information needs and then used the system to search, browse the retrieved content, and read the generated reports. To ensure the representativeness and openness of the tasks, a heuristic task generation approach was adopted. The researchers provided nine predefined topic categories, covering: practical guides, career development, product/service comparison, health and psychology, academic and knowledge learning, life experience sharing, consumption and word-of-mouth, travel and location exploration, and social issues and value judgments. Participants could select topics according to their personal interests and freely generate specific queries. To maintain task complexity and exploratory depth, overly simple or factual questions (e.g., ``What is the weather today?'' or ``Who won the 2025 Nobel Prize in Physics?'') were excluded. After the experiment, the evaluations were conducted through a mixed-method approach that combined questionnaires and interviews. Participants rated their perceptions and interaction experiences in information seeking on a five-point Likert scale.

\subsubsection{Data Collection and Analysis}

We adopted a mixed-methods approach that integrated both qualitative and quantitative analyses. Our goal was not only to examine whether social cues led to better or worse outcomes, but also to understand how users perceived these cues and why such changes occurred. Considering the substantial differences in motivation and experience across participant roles, quantitative measures alone were insufficient to capture the nuanced mechanisms behind user behaviors. Inspired by prior studies \cite{yang2025socialmind, qian2025exploring, ramjee2025cataractbot}, we therefore combined qualitative insights with quantitative evaluation in our data collection and analysis.

For the quantitative analysis, we focused on participants’ evaluations of both the perceived quality of information seeking results and their experience.
Specifically, at the perception level, we examined Perceived Usefulness, Trustworthiness \cite{jiang2024into, prabha2007enough, yamamoto2018exploring}.
At the interaction experience level, we focused on Serendipity, Willingness to Use, Sense of Direction, and User Control \cite{dork2011information, prabha2007enough, wilson1999models}.
The detailed quantitative dimensions are summarized in Table~\ref{tab:measurement_dimensions}:
\begin{table}[htbp]
\centering
\small
\caption{Evaluation dimensions of user perception and experience.}
\renewcommand{\arraystretch}{1.2}
\begin{tabular}{lp{11cm}}
\toprule
\textbf{Dimension} & \textbf{Description} \\
\midrule
Perceived Usefulness & The information provided by the system effectively helps me accomplish my current tasks or answer my questions. \\ \addlinespace
Perceived Trustworthiness & I believe that the search results presented by the system are authentic and reliable. \\ \addlinespace
Willingness to Use & Considering both the interaction process and the information obtained, I am willing to continue using this system for information seeking. \\ \addlinespace
User Control & I feel that I can actively control the direction of search and filtering without being constrained or dominated by the system. \\ \addlinespace
Sense of Direction & Throughout the information seeking process, I am always aware of what I am looking for and do not feel lost or confused (e.g., losing track of the original goal or uncertain about the next steps in the search). \\ \addlinespace
Serendipity & During the use of this system, I am able to discover unexpected yet valuable new information. \\
\bottomrule
\end{tabular}
\label{tab:measurement_dimensions}
\end{table}
    
    

    
    
We computed the mean and variance of each metric, visualized their data distributions, and performed significance tests to assess differences between the experimental and control groups.

For the qualitative data, we collected users’ cognitive and behavioral responses through a combination of think-aloud protocols and semi-structured interviews. Participants were encouraged to verbalize their thoughts while rating, allowing us to capture their reasoning processes. In addition, we observed and recorded users’ interaction behaviors during information seeking with the system. The collected data were analyzed using an inductive thematic analysis approach \cite{braun2006using}, aiming to identify how the integration of social cues influenced participants’ information seeking processes. For example, when a participant mentioned being ``frustrated that the model couldn’t distinguish common household ingredients,'' we coded such statements as ``difficulty in accurately interpreting social cue conditions'' and ``negative emotions caused by misinterpretation of social cues.'' This open coding process ensured that the analysis remained grounded in participants’ own language and reflections, encompassing both objective observations and subjective experiences. 

After the initial coding, the researchers collaboratively grouped related codes into broader and more abstract themes. For instance, ``discovering new information through consensus in the comment section'' and ``spatial associations between content inspiring further exploration'' were merged into the theme ``Information Serendipity.''
Similarly, codes such as ``limited model understanding of complex social relationships,'' ``insufficient social commonsense knowledge,'' and ``limited understanding of regional cultures'' were categorized under the theme ``Limited Social Knowledge Representation of the Model.''
This iterative process involved multiple rounds of cross-comparison to ensure that each theme accurately reflected participants’ experiences and the nuances of their interactions with the system.

Finally, through the integration of qualitative analysis, quantitative analysis, and cross-validation, we identified four major themes, revealing both the potential values and limitations of integrating social cues into LLM-based search: (1) Enhancing User Perception of Information Need Fulfillment,(2) Enhancement in User Experience,(3) Enhancement in User Reflection, (4) Facing Challenges in Aligning Users within Complex Social Cue Queries.

\begin{table}[htbp]
\centering
\caption{Comparison of metrics between Group E and Group C. $\Delta$ represents the mean difference (E–C).}
\renewcommand{\arraystretch}{1.2}
\begin{tabular}{llcccccc}
\toprule
\textbf{Category} & \textbf{Dimension} & \textbf{C$_{Mean}$} & \textbf{C$_{SD}$} &
\textbf{E$_{Mean}$} & \textbf{E$_{SD}$} & \textbf{$\Delta$} & \textbf{p-value} \\
\midrule
\multirow{2}{*}{\textit{Perception}} 
 & trustworthiness & 3.50 & 0.83 & 4.55 & 0.60 & +1.05 & 0.0002 \\
 & usefulness      & 2.75 & 0.85 & 3.85 & 1.09 & +1.10 & 0.0011 \\
\midrule
\multirow{4}{*}{\textit{Experience}} 
 & user control   & 2.60 & 0.88 & 4.20 & 0.52 & +1.60 & <0.0001 \\
 & sense of direction       & 3.05 & 0.94 & 4.50 & 0.61 & +1.45 & <0.0001 \\
 & serendipity     & 2.20 & 0.83 & 4.00 & 0.86 & +1.80 & <0.0001 \\
 & willingness to use     & 2.55 & 0.83 & 4.20 & 0.62 & +1.65 & <0.0001 \\
\bottomrule
\end{tabular}
\label{tab:comparison_statistics}
\end{table}

\subsection{F1: Enhancing User Perception of Information Need Fulfillment}
\label{sec:f1}
Table~\ref{tab:comparison_statistics} presents the detailed comparison of user evaluations, and Fig.~\ref{fig:compare_box} shows the rating distributions. 
The experimental group reported higher perceived usefulness (+1.10) and trustworthiness (+1.05) than the control group. 
The distribution of trustworthiness ratings was more concentrated than that of usefulness, reflecting greater agreement among participants on the reliability of the results, while perceptions of usefulness varied more across individuals. 
Our qualitative analysis further explains the underlying reasons for these improvements.

\subsubsection{Enhancing Perceived Usefulness}
 During the search process, users integrated various social cues to judge and refine their search criteria, allowing LLMs to more effectively filter content and generate results better aligned with their needs. Participants typically considered multiple dimensions, including content style, comments and author identity. 
 E7 described using stylistic cues to infer content reliability:\interviewquote{``I judge based on the content style. For example, I can exclude posts with overly intimate terms like ‘Baozimen’ (Dear ones) or ‘Jiarenmen’ (Family) when searching for authoritative information.''}. E3 further emphasized the diagnostic value of comments as a form of collective opinion rather than mere engagement metrics:
\interviewquote{``Comments feel more genuine than the number of likes because they require more effort from people to write. The system can effectively use these viewpoints to enhance response quality.''} E8 mentioned that knowing the poster’s social identity or background helped them better interpret the credibility and applicability of posts:\interviewquote{``For open-ended questions, what I really want to know is the identity of the person involved… I can filter for experiences from people whose situations are similar to mine.''} In contrast, participants in the control group also expressed a desire for social cue–based filtering, but their system could not adequately support it. C8 reported frustration when the retrieved results contained clickbait or marketing-oriented posts that the model failed to filter out:
\interviewquote{``…a title like ‘Breaking News! Salary Explosion!’—most likely an exaggerated promotional post. What I actually wanted were genuine personal experiences.''} C5 also emphasized the importance of assessing comment feedback when searching for practical guides, noting that they \interviewquote{``hope to hide the ones with many negative comments, but the system doesn’t support...''}

During the result filtering phase, users generally believed that the integration of social cues helped them more efficiently exclude irrelevant or unanticipated content. They noted that the LLM could often interpret their cue conditions correctly and filter results accordingly. For example, E3 mentioned that seeing how the system explicitly explained the cue–result alignment increased their confidence in the model’s understanding:
\interviewquote{``When I saw the explanation, I felt that it understood correctly.''}

Moreover, when the LLM generated the final result report based on the filtered posts, several participants noted that social cues influenced not only their understanding of the results but also how the system structured the information. They observed that after expressing social cues, the system could better organize and prioritize content in line with their personal interests, producing reports that were easier to interpret and more relevant. E1 described that the system’s organization became noticeably clearer after incorporating cues:
\interviewquote{``After I expressed the social cues, the system reflected my requirements more clearly in the organization of the report.''}

\subsubsection{Enhancing Perceived Trustworthiness}

The integration of social cues substantially enhanced users’ perceived trustworthiness of the search results. Participants reported that when the system matched results with their social cue conditions, it explicitly presented the rationale behind each match. This explanatory feedback helped users understand the system’s judgment process and evaluate whether its reasoning aligned with their expectations, thereby strengthening their trust in the outcomes. Several participants mentioned that such transparency made it clear why certain posts were included or excluded. E1 observed that after reviewing the model’s explanation, it accurately reflected their intent and filtering logic:
\interviewquote{``After reviewing the model’s explanation, I found that it accurately translated my requirements.''} Others appreciated that the system provided graded explanations rather than simple scores, offering both numerical and textual rationales. E8 noted that this approach improved confidence in the model’s judgment:\interviewquote{``...when the explanation made sense, it helped me build trust in the results.''}

Beyond the trust built through explainable cue matching, participants mentioned that social reference—seeing results that reflected their own social conditions or identities—also strengthened their confidence in the credibility and applicability of the information. They perceived the retrieved posts as being more aligned with their own situations, which provided an explicit social anchor for judging reliability.E1 noted that content based on real personal experiences from socially similar users felt particularly trustworthy:
\interviewquote{``Personal real-life experiences are more clearly reflected. A group of people whose location, goals, or identities are similar to mine can provide experiences and advice that are truly helpful.''} E3 emphasized that shared experiences from peers carried more weight than institutional sources:
\interviewquote{``When I am a consumer, I tend to trust other consumers rather than businesses. If I see that others have sought similar information or shared their experiences, I am much more likely to trust that content.''} E4 added that perceiving others’ summaries or firsthand usage increased the perceived reliability of results:
\interviewquote{``It feels like they have actually used it or provided some summaries, so selecting from these summaries seems more reliable.''} Similarly, E7 mentioned that results from users with similar backgrounds reinforced the sense of credibility:
\interviewquote{``if someone with a similar identity to mine, also a computer science major, is searching for similar job posts, then the information from peers is more trustworthy.''}

\begin{figure}[h!]
  \centering
  \includegraphics[width=0.92\textwidth]{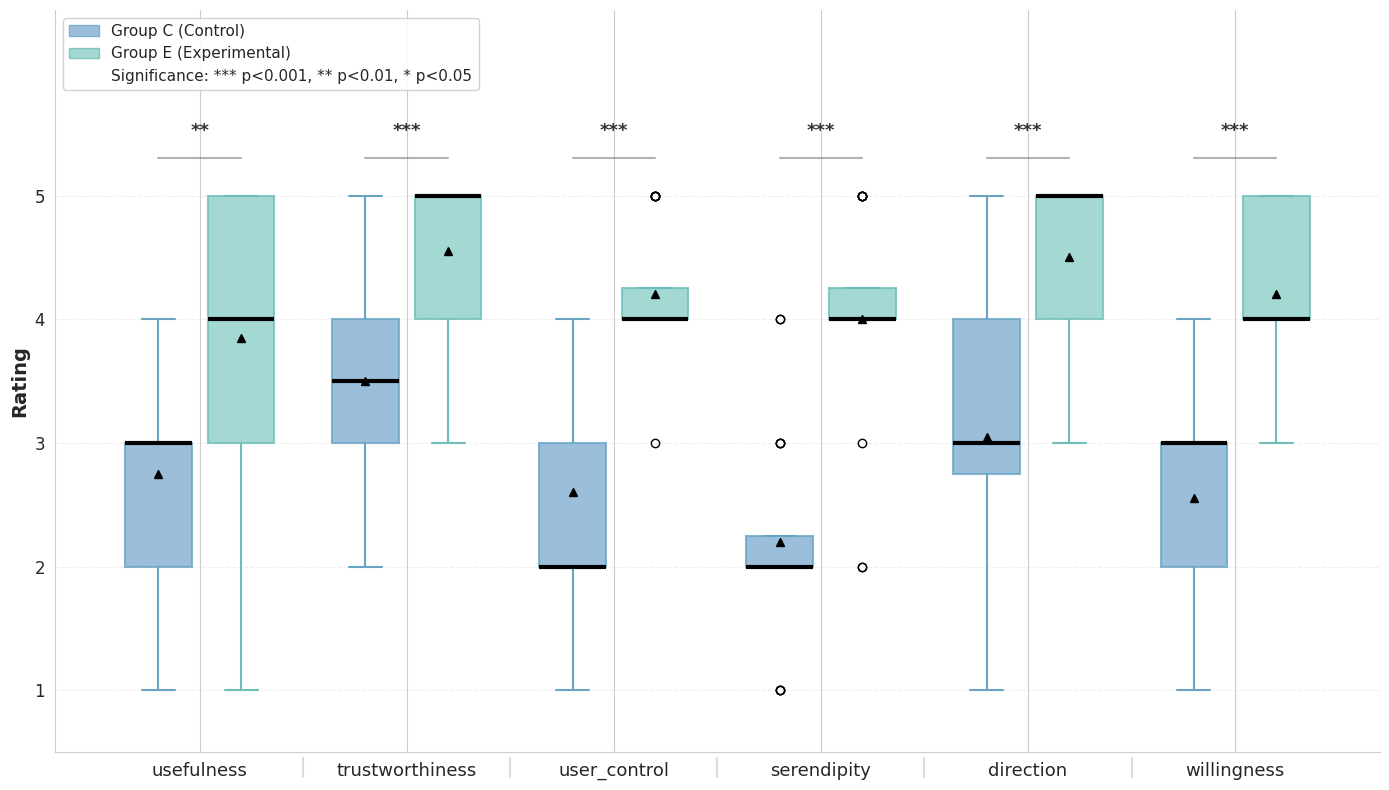}
  \caption{Distributions of user ratings across evaluation dimensions between Group E and Group C.}
  \label{fig:compare_box}
\end{figure}



\subsection{F2: Enhancement in User Experience}
\label{sec:f2}
The integration of social cues not only enhanced users' perception that their information needs were better satisfied, but also improved their subjective experience during the information-seeking process. 
As shown in Table~\ref{tab:comparison_statistics} and Fig.~\ref{fig:compare_box}, the experimental group achieved consistently higher ratings across all experiential dimensions, including user control (+1.60), sense of direction (+1.45), serendipity (+1.80), and willingness to use (+1.65). 
According to the boxplot distributions, in all four dimensions, the lower quartile (25\%) of the control group fell below a rating of 3, indicating generally negative experiences, whereas in the experimental group, all lower quartiles rose above 3. Our qualitative analysis further explains the specific patterns behind these improvements.

\subsubsection{Enhancing User Control}

Participants in the experimental group demonstrated a stronger ability to control during information seeking, flexibly adjusting social cues based on their own characteristics and preferences to obtain results better aligned with their needs.
Unlike the passive recommendation mechanisms of traditional platforms, this system allowed users to directly intervene in the model’s decision-making process through interactive engagement, enabling them to participate in content filtering and generation at an early stage of the search.
As E1 described, being able to view and respond to the model’s intermediate reasoning empowered them to shape the outcome rather than passively waiting:
\interviewquote{``I could agree or disagree with them, and the model would adjust accordingly.''}

This form of pre-intervention transformed users from passive reviewers into active participants in the search process. Rather than waiting for system outputs, they guided the model’s matching logic through personalized feedback, gaining a stronger sense of control.
E3 described that the experience was fundamentally different from their usual behavior on RedNote’s original LLM-based search:
\interviewquote{``I usually get an overview first and then browse individual posts. Here, it’s the opposite—I read individual posts first and then summarize what I agree with. I feel a stronger sense of control because I can check and adjust throughout the process, and the final report reflects my own choices.''} In contrast, C10 highlighted the limitations of conventional LLM-based search, noting the lack of fine-grained control: \interviewquote{``When I search for posts myself, I can refresh or re-select tags to refine what I see. But with the model-based search, it often retrieves a large batch of posts—including many I don’t want—and even after restarting, those unwanted ones may still reappear.''}

Through continuous interaction with the model, some users experienced a shift from viewing it as a mechanical summarizer to recognizing it as an intelligent collaborator capable of joint reasoning and co-creation. E2 described this change: \interviewquote{``At first, I saw it as just a summarizer, but later I realized that I could take control of it. It now feels like a collaborator that understands my intent and organizes information accordingly.''} This cognitive shift reflects a reconfiguration of user control in information seeking, where users evolve from mere operators of a search system to co-creators who collaboratively shape outcomes with the model, resulting in a more empowered and participatory search experience.

\subsubsection{Enhancing Serendipity}
Although our initial intention in introducing social cues was to help users filter and interpret information, interestingly, we found that social cues also opened up opportunities for users to explore new information—and in some cases, even became a goal of information seeking themselves. By revealing cues such as authors’ geographical locations, the system helped users notice previously overlooked relationships and make unexpected yet valuable discoveries. For instance, E7 searched for ``how to make wonton soup'' and appreciated that the displayed locations prompted a regional reinterpretation of the recipes:\interviewquote{``If it hadn’t listed the locations, I wouldn’t have thought from that perspective. In southern wonton soup, they tend to use more lard, while in the north, it’s more about sesame oil.''}

By leveraging social cues, the system helped users uncover details or relationships they had not previously noticed, leading to unexpected yet valuable discoveries. One participant, for example, searched for ``how to make wonton soup'' and noticed that the system displayed social cues indicating the authors' geographical locations. This prompted him to reinterpret the recipes from a geographical perspective, leading to an interesting discovery of the differences between northern and southern variations of the dish.
\interviewquote{E7: ``If it hadn't listed the locations, I wouldn't have thought from that perspective. That information was quite valuable — in southern wonton soup, they tend to use more lard, while in the north, it's more about sesame oil.''}

The participant noted that social cues helped them move beyond the traditional semantic retrieval paradigm, enabling a reorganization of the information space through social contexts.
E8 explained that this shift provided greater flexibility in exploration:
\interviewquote{``Previously, whether embedding-based or keyword-based, search was always about the content itself. With social cues, I can loosen content constraints and use them to define a more meaningful boundary for exploration.''}

Participants observed that social cues helped them uncover consensus patterns beyond individual posts, revealing broader community dynamics. The structure of comments and feedback was seen not as supplementary information but as a key indicator of collective opinions and social sentiment.
E10 noted that \interviewquote{``the consensus reflected in the comments can actually provide highly valuable information.''} In this sense, the novelty brought by social cues lies not only in the unexpectedness of new information but also in the reframing of social dynamics. By observing how others respond, resonate, or disagree, users developed a higher-level understanding — shifting from individual information acquisition to a more holistic perception of community cognition and social structures.

\subsubsection{Enhancing Sense of Direction}

Participants described social cues as navigational aids that provided a stable logical thread during information exploration. Compared with relying solely on keyword or semantic matching, social cues helped them establish clearer reference points within the complex information space. E1 referred to social cues as ``a logical thread that connects everything together,'' while E10 described ``searching along two parallel lines—one for content and another for social cues.'' As E2 noted, social cues also made their intentions more explicit and traceable throughout the search process, ``constantly reminding me of what I’m looking for.'' This dual-track process enabled users to engage in simultaneous searching and reasoning across both content and social dimensions, fostering a clearer sense of direction before the results are generated.

For goal-oriented search tasks, this sense of direction became particularly salient. Social cues acted as attentional scaffolds, helping users efficiently filter out irrelevant or distracting content—such as advertisements—and maintain cognitive focus.
E3 explained that these cues reinforced the intended context during academic exploration:
\interviewquote{``When there are advertisements, the system marks them as ads, so I naturally ignore them. And when it shows who’s involved—like which university or lab—they help me stay within an academic search context.''} Such seeking reminders were also task-dependent, as E9 noted:\interviewquote{``When I’m just browsing for fun, I don’t need... But when I have a clear goal, this matching process becomes essential for keeping that direction.''}

When evaluating the generated results, users relied on social cues to assess the credibility and richness of reports. Variations in social attributes—such as follower counts, comment hierarchies, and user types—helped them quickly judge the authenticity and usefulness of aggregated content.
E1 found that these cues revealed patterns of genuine participation:
\interviewquote{``I wanted to read authentic personal experiences, and I noticed that many genuine users actually have very few followers.''}
E6 appreciated that such information was centrally displayed, eliminating the need to check each user profile individually:
\interviewquote{``In the original system, I had to click into each user’s profile to see their field or follower count. Here, all those details appear in one place.''}  In contrast, C9 mentioned that ``when I check an author’s profile, I get distracted by the images in past posts,'' noting a need for help to stay focused on the search.

\subsubsection{Enhancing User Willingness to Use}

In LLM-based information retrieval systems, users who employed social cues generally showed a higher willingness to use the system. Although incorporating social cues introduced some cognitive load during expression and use, participants found it acceptable and even worthwhile. E1 pointed out that,  \interviewquote{``as a social being, I feel this is simply instinctive,''} further explaining that: \interviewquote{``When searching on social content platforms, I naturally process such cues anyway, and the system’s visualization of their relationships makes the effort worthwhile.''}. Similarly E2 expressed \interviewquote{ ``Overall, it felt very effortless, almost instinctual. Actually, on the platform, I often spontaneously use social cues myself.''}

Besides, participants appreciated using natural language as a medium for expressing social cues, valuing its flexibility and the model’s ability to handle imprecise or loosely defined descriptions. E4 explained:
\interviewquote{``Since I can search directly using natural language, there are fewer constraints. Unlike before, when I had to think of precise keywords and worry about typos, the system now provides explanations that I can refine or correct as needed.''}

Some participants noted that social cues created a ``mental matching pattern'' even before issuing a query, helping them quickly judge whether the system understood their intent and thus reducing the verification burden.
E2 explained:
\interviewquote{``Social cue matches are generally more apparent than content matches—you can immediately tell if the system understands what I mean.''}
This expectation effect enabled users to focus more rapidly on key posts, minimizing the time and effort spent on repeated verification.

\subsection{F3: Enhancement in User Reflection}
\label{sec:f3}

Based on observations of user interactions and analyses of interviews, we found that social cues also prompted users to reflect on their own information seeking behaviors. Users began to reassess the validity of information sources and, through the process of expressing and utilizing social cues, continuously refined their information needs and evaluation criteria.

\subsubsection{Validity of Information Sources}
During information seeking with social cues, some participants began to reflect on the suitability of the platform itself for their information needs.
For example, E5 searched for civil service exam strategies and set strict social cues—targeting posts by verified civil servants or senior peers with formal, detailed, and advertisement-free content. When many retrieved posts were labeled non-compliant by the model, the participant recognized a mismatch between the platform’s ecosystem and the task objectives.
Similarly, E4 described this experience metaphorically:
\interviewquote{``It’s just like fishing—social cues help me sense the depth and flow of the water, letting me know whether the fish I’m looking for are here. If not, I know it’s time to move on.''}

\subsubsection{Expression of Needs}
Participants noted that considering which social cues to include in their queries prompted them to revisit and refine their problem definitions.
E2 explained that defining social constraints on content, authors, and feedback led to a clearer understanding of information expectations:
\interviewquote{``Thinking about how to constrain content and authors made me reflect on what I actually expected from the information.''}

Several participants reflected on how social cues prompted them to reconsider the priority of different conditions in their queries. Some realized the need to distinguish between essential and optional conditions.
E9 noted that strict cue settings reduced the number of matching posts, revealing that even partially matched content could still be useful:
\interviewquote{``After applying my social cues, I found that strictly matching posts decreased, but some mismatched ones were still valuable. It made me wonder whether I should relax certain conditions or let the model know which cues weren’t strict requirements.''}
This process also encouraged users to reflect on their deeper information goals. As E8 explained:
\interviewquote{``Initially, I required first-hand experiences of moving from industry to academia, but I later found second-hand accounts equally helpful. It made me realize that what I really want is information I can compare to my own situation.''}

\subsection{F4: Facing Challenges in Aligning Users within Complex Social-Cue Queries}

\subsubsection{Deficient Social Knowledge}

The LLM’s understanding of social knowledge remains limited, particularly in recognizing community dynamics and social relationships.
E2 highlighted that this gap can lead to misinterpretations of nuanced group contexts:
\interviewquote{``LLMs lack understanding of certain social dynamics. For example, within an idol group, different fan subcommunities exist—some fans like members X and Y but dislike Z. If the model ignores this, it may recommend unwanted content.''}

When social cues involve common sense or situational knowledge, the model’s comprehension becomes even more constrained. It often struggles to translate users’ natural language conditions into actionable filtering criteria.
E6 illustrated this limitation:
\interviewquote{``When I searched for wonton soup recipes using only common household seasonings, the model seemed unsure what ‘common’ meant and included ingredients that most people don’t usually have.''}

Inaccuracies in the model’s understanding of geographical and cultural contexts further limited the effectiveness of social cue matching.
E10 noted that the model often misinterpreted contextual nuances:
\interviewquote{``The model labeled a post as a tourist guide, but it was actually a local weekend trip from Kunming to Yuxi—two places very close to each other. It still can’t distinguish such subtle geographical differences.''}
Similarly, E1 observed that while some social cues were correctly recognized, the model struggled to interpret how they should manifest in content:
\interviewquote{``When I specify that a post should contain ‘enough details,’ that standard exists only in my mind. The model needs to learn from my feedback and behavior to better align with my expectations.''}

\subsubsection{Overprocessing of Social Cues}

Some participants observed that the model sometimes overemphasized certain social cues, resulting in biased or overly strict matching. Even when a single cue dimension was correctly interpreted, multi-criteria queries combining source, content, and comment features often triggered excessive filtering and reduced recall.
E8 noted that this strictness was unnecessary:
\interviewquote{``The recall was reduced without reason. We usually prioritize recall over precision because the goal is to retrieve relevant items. The post itself was fine, but one comment misled the model, and it got filtered out.''}

Furthermore, when generating results, the model tended to amplify dominant social cues, such as those from professional creators, while overlooking content from ordinary users. This bias risks reducing viewpoint diversity and leading to homogenized summaries over time.
E7 expressed concern about this imbalance:
\interviewquote{``The current results obscure the diversity of social cues. Although I can later instruct the model to include specific cues, I’d prefer to set this proactively when generating reports—for example, to ensure a diversity of perspectives rather than everyone repeating the same view.''}

%% file: 5_Discussion.tex
\section{Discussion}

\subsection{Social Cues as a Lens: Understanding Users’ Needs for Agency}

In the current LLM-based search, the system can automatically summarize large amounts of information and generate responses that seemingly reduce users’ cognitive load. However, this highly automated functionality often undermines users’ agency\cite{schroeder2025forage, cheng2025ragtrace}, which in information seeking can be understood as the extent to which users actively initiate actions and align them with their own values and identities\cite{bennett2023understand}. During automated integration and generation, the sources and reasoning logic are often obscured, creating a black-box effect that makes it difficult for users to understand, intervene in, or control the process. This opacity disrupts users’ original cognitive pathways of information processing, introducing new cognitive load when interpreting model-generated summaries, and shifting users from proactive explorers to passive recipients. Yet the nature of human information seeking is inherently iterative and incremental—a gradual loop of hypothesis formation, exploration, and evaluation—where each step must remain within the seeker’s cognitive capacity \cite{wilson1999models}.


Our findings for RQ1 (how users expect to use social cues) reveal that users exhibit a desire to engage the seeking process through configuring social cues such as the author profile, content style, and comment context, thus intervening query conditions, calibrating matching results, and exploring alternative results or reports. Our findings for RQ2 (how social cues take effect) reveal that users leverage social cues to regain their lost agency in filtering and making sense of retrieved information. Beyond this recovery, we also observed the emergence of new forms of user agency in three aspects.

\textbf{User agency in clarifying information needs.}When users articulate their queries through social cues, such as considering whose post they want to see, what type of post is appropriate, and how readers respond, they are able to formulate clearer representations of their message-level queries and uncover the underlying essence of their information needs. In other words, expressing search intent through social cues provides an opportunity for reflective clarification, allowing users to better understand what they truly seek.

\textbf{User agency in maintaining the seeking process.}
The integration of social cues provides users with a clearer sense of direction throughout the search process. Quantitatively, users in the social-cue condition reported a 1.45-point higher sense-of-direction rating on average compared with the control group. Qualitatively, participants explained that matching with social cues served as an ongoing reminder or anchor, helping them sustain attention, maintain task focus, and preserve cognitive continuity during information seeking.

\textbf{User agency in exploring information at higher dimensions.}
With the support of LLMs’ text-processing capabilities, users can engage in higher-order sensemaking across multiple layers of social information. By leveraging social cues, they construct multidimensional information schemas, such as identifying consensus within specific author communities, synthesizing insights from multiple comment threads, and comparing content under different social contexts. These practices demonstrate how social cues extend user agency from individual interpretation to collective-level exploration and synthesis.

A potential explanation for these newly discovered forms of agency lies in the way users’ implicit cognitive processing of social cues \cite{winter2016selective} becomes increasingly explicit and reflective during interactions with LLM-based systems, enabling them to develop greater awareness of both the cues themselves and the relationships among them.

\subsection{Advantages and Weaknesses of LLM-based Information Seeking with Social Cues}
\textbf{The integration of social cues into LLM-based search enhances users’ perceived need fulfillment, experiential quality, and reflective capacity throughout the information seeking process.} For enhancing the user perception level, social cues improve the interpretability and credibility of results, enabling users to understand the logic of result generation, thereby increasing their trust and acceptance of model outputs.  It is consistent with "social cues promote social cognition and trust" \cite{feine2019taxonomy} and also corresponds to the socialized retrieval augmented generation effect \cite{wang2025social}, that is,  the system can improve the relevance and transparency of recommendations by leveraging previous interactions or group feedback. At the level of enhancing user experience, social cues give users a stronger sense of active control and exploration, enabling users to transform from passive result receivers to active process participants\cite{liu2022building,li2025demod}. This reflects the mixed- initiative mechanism in HCI field \cite{mei2025interquest}, that is, system and user jointly shape the retrieval target through the sharing of social cues. Similarly, \cite{hu2013whoo} and \cite{de2014seeking} demonstrate that in online environment, social cues can stimulate users' social motivation and belonging, enhancing their willingness to continuously participate community. Our experiments also prove that social cues can trigger this socially-driven exploration with information flaneur style in LLM-based search \cite{dork2011information}. At the level of promoting user reflection, social cues prompt users to verify information sources and expression needs, improving the search quality and providing  opportunities to cultivate information literacy. This aligns with the viewpoint \cite{neshaei2025user}, the system should prompt users to reflect rather than replace their thinking process. Users can identify the social structure and value orientation behind the information by leveraging social cues, which also echoes the cyclic verification-integration-reflection dynamics proposed in the Personal Information Ecosystem \cite{milton2024seeking}.

\textbf{LLM-based search with social cues still faces key weaknesses in social knowledge reserves and multi-dimensional cue balancing.}
Firstly, LLMs' abilities in understanding complex relationships, cultural contexts, and social common sense are limited, leading to deviations in social cue matching. This issue aligns with that the LLM's understanding deficiency in group social contexts \cite{wang2025social}, indicating that models' knowledge modeling still remains at the shallow semantic matching stage and lacks the capture of interaction dynamics. Second, when dealing with multiple conditions, the model tends to overemphasize one dimension, resulting in homogenization of results and weakening the diversity that social cues should offer. Specifically, during the generation, LLMs tend to amplify the most significant signals and neglect the balance among different cues. Since these social cues covering various types such as identity, group feedback, and interaction behaviors, the model may be potentially guided by the feature with the highest statistical probability (for example, a majority of a certain occupation in search results), thereby introducing bias towards certain cues. Similar biases are also observed in information retrieval and recommendation system research\cite{liu2025unbiased, shu2024rah}. In information retrieval, models often favor high-frequency words or features in prominent positions \cite{sharma2024generative}, while recommendation algorithms are prone to popularity bias and exposure amplification \cite{chen2023bias}. Therefore, explicit weight allocation or multi-objective optimization strategies should be introduced in the future model design during the social cue integration stage, alleviating the problems of significant cue amplification and diversity loss.

\subsection{Design Implications for Future Information Seeking Systems}
Building on the above findings and prior studies, future LLM-based information seeking systems should make dual improvements at both the model level and the system level. 

\textbf{At the model level, efforts should be made to enhance the integration of social knowledge and diverse contextual reasoning. } Empirical studies reveal gaps in LLMs' understanding of social knowledge \cite{wang2025social}, such as suboptimal performance on social norms tasks (e.g., achieving only moderate accuracy across benchmarks like SocKET \cite{choi2023llms}). The gap has also been underscored in recent studies on LLM-based voice assistants \cite{qian2025exploring}, highlighting the importance of strengthening the training of foundational models to better capture and reason about social context. Future models should enhance the understanding of implicit social semantics by introducing social knowledge graphs and cross-modal social data in the training process. 
For instance, through graph injection techniques\cite{pan2024unifying}, models can learn implicit relationships from social network data, which in turn improves the understanding of subtle cues like author identity or the overall interaction atmosphere within conversations. Additionally, cross-modal training can integrate text-image-audio data (e.g., from social media posts) using self-supervised learning to align modalities, ultimately achieving more robust recognition of social cues that span multiple forms of input \cite{gao2025large, hallgarten2025routellm}. 
Besides,  the"taxonomy of social cues" concept can be used as a guidance framework to model social context, encompassing three feature representations including the individual, group, and culture. The individual representation encodes linguistic style and personality cues; the group representation identifies social roles and interaction patterns; and the cultural representation reflects community norms and pragmatic conventions. During reasoning, models should dynamically balance these representations to avoid overfitting certain significant signals, thereby enhancing contextual sensitivity and social coherence in generated outputs.

\textbf{At the system level, future search interfaces should enable negotiable cue configuration and transparent feedback mechanisms. } Inspired by mixed-initiative design principles \cite{mei2025interquest, neshaei2025user, 10.1145/3432193}, systems can allow users to personalize and iteratively refine their social cue preferences, forming adaptive cue templates over time.
Moreover, based on the thought of explainable retrieval–generation \cite{cheng2025ragtrace}, systems should visualize model reasoning via cue cards or cue trace maps to display referenced social cues and relevant weights to allow users to click and intervene in the matching logic, fostering user understanding and intervention in time and further enhancing the co-search experience.
Additionally, systems should provide recall–precision balancing controls, such as integrated slider controls,
 enabling users to switch the mode according to the task goals (exploratory, confirmatory, or comparative), thereby dynamically optimizing search modes. This design aligns with the cost–benefit balance though in the information foraging theory \cite{pirolli1995information} and also resonates with recent research emphasizing user agency and exploratory engagement \cite{bennett2023understand}.

%% file: 6_Limitation.tex
\section{Limitations}
Although the study reveals the actual impact of social cues in LLM-based search, several limitations remain regarding the research scope and experimental conditions. (1) The study was conducted on RedNote, whose distinctive culture and user base may influence the results. While the platform provides rich and authentic social cues, the findings may not generalize to other platforms (e.g., Reddit, Zhihu, Twitter). Moreover, participants were digitally literate and active users, possibly more familiar with LLM-based search and social cues than average users, which may affect generalizability. (2) The experiments focused on open-ended information seeking tasks, excluding factual retrieval or multi-turn collaborative scenarios. The effectiveness of social cues may vary with task goals (e.g., decision support vs. knowledge acquisition). Future work should examine their adaptability across different task types. (3) The system was tested within a Chinese content platform, where cue extraction and interpretation rely on specific linguistic and cultural conventions. Since social expressions and norms differ across languages and cultures, the system’s cross-cultural adaptability requires further validation.

%% file: 7_Conclusion.tex
\section{Conclusion}
This study systematically examined how social cues operate in LLM-based information seeking. Through design workshops and empirical studies with our prototype \ours, we revealed how users actively utilize and configure social cues during LLM-based search, and how these cues shape their judgment, sense of direction, and sense-making across different stages (pre-, in-, and post-search). The results show that social cues not only play a supportive role in information filtering and comprehension but also promote users’ active participation and cognitive reflection, transforming the information seeking process from passive reception to proactive collaboration. The integration of social cues significantly enhances users’ trust in the results, sense of direction, and information serendipity. However, the study also finds that under complex social contexts, large language models still have limitations in possessing sufficient social knowledge, interpreting cross-culture, and balancing multi-dimensional conditions, which may lead to cue overfitting or semantic bias. Future research could strengthen models’ representation and reasoning abilities for social and cultural contexts at the model level, and explore how to balance transparency, controllability, and automation at the system design level, to support a more socially aware and human–AI co-engaged search experience.